\documentclass[twocolumn,twoside,nofootinbib,showpacs,prd,aps,tightenlines,10pt]{revtex4-1}

\usepackage{amsfonts,amsmath,bm,mathrsfs}
\usepackage{graphicx}

{\count255=\time\divide\count255 by 60 \xdef\hourmin{\number\count255}
  \multiply\count255 by-60\advance\count255 by\time
  \xdef\hourmin{\hourmin:\ifnum\count255<10 0\fi\the\count255}}

\def\nn{\nonumber \\ }
\def\rd{{\rm d}}

\def\vev#1{\left\langle #1 \right\rangle }

\def\hsix{ \mathcal{H}^{(6)}}
\def\ks{ \mathfrak{m}_S}

\begin{document}

\title{An Exactly Solvable Model for Dimension Six Higgs Operators and $h \to \gamma \gamma$}

\author{Aneesh V.~Manohar}

\affiliation{Department of Physics, University of California at San Diego,
  La Jolla, CA 92093\vspace{4pt} }

\begin{abstract}
An exactly solvable large $N$ model is constructed which reduces at low energies to the Standard Model plus the dimension six Higgs-gauge operators
$g_1^2 H^\dagger  H  B_{\mu \nu} B^{\mu  \nu}$, $g_2^2 H^\dagger   H  W^a_{\mu \nu} W^{a\mu  \nu}$, 
$g_1 g_2 H^\dagger  \tau^a  H  W^a_{\mu  \nu} B^{\mu \nu}$, and $\epsilon^{abc}  W^a_{\mu}{}^\nu W^b_{\nu}{}^\rho W^c_{\rho}{}^\mu$. All other dimension six operators are suppressed by powers of $1/N$.
The Higgs-gauge operators lead to deviations from the Standard Model $h \to \gamma\gamma$ and $h \to \gamma Z$ rates. A simple variant of the model can be used to also generate the Higgs-gluon operator $g_3^2 H^\dagger   H  G^A_{\mu \nu} G^{A\mu  \nu}$ which contributes to the Higgs production rate via gluon fusion.
\end{abstract}

\maketitle

A scalar boson with a mass of $\sim 125$\,GeV has recently been discovered at the LHC, and it is important to study its properties in a model-independent way. The Standard Model provides a good description of the LHC data so far, with no evidence for any new particles with masses below $\sim 1$\,TeV. A widely used starting point for analysis is to assume that the theory at $125$\,GeV is the Standard Model including a fundamental scalar doublet, and all new physics effects are characterized by higher dimension operators involving Standard Model fields.

A recent paper~\cite{Grojean:2013kd} considered the impact of dimension-six operators on the Higgs decay rate.
The theory considered was the Standard Model plus the dimension six Hamiltonian
\begin{align}
\hsix = -\mathcal{L}^{(6)} &= c_{G}  \mathcal{O}_{G}+c_{B}  \mathcal{O}_{B} + c_{W}  \mathcal{O}_{W} + c_{W\!B}  \mathcal{O}_{W\!B}\label{h6}
\end{align}
generated by new physics at some scale $\Lambda$, where
\begin{align}
\mathcal{O}_{G} &=  \frac{g_3^2}{2  \Lambda^2}  H^\dagger   H  G_{\mu \nu}^A G^{A \mu  \nu},  
&\mathcal{O}_{B} &=  \frac{g_1^2}{2  \Lambda^2}  H^\dagger   H  B_{\mu \nu} B^{\mu  \nu},   \nn
\mathcal{O}_{W} &=  \frac{g_2^2}{2  \Lambda^2}  H^\dagger   H  W^a_{\mu \nu} W^{a\mu  \nu},  
& \mathcal{O}_{W\!B} &=  \frac{g_1  g_2}{2  \Lambda^2}  H^\dagger  \tau^a  H  W^a_{\mu  \nu} B^{\mu \nu} .
\label{ops}
\end{align}
using the notation of 
Refs.~\cite{Manohar:2006gz,Manohar:2006ga}.  These operators give amplitudes which can interfere constructively or destructively with the Standard Model amplitudes for $gg \to h$, $h \to \gamma \gamma$, $h \to  Z\gamma$, etc.. The phenomenology of the operators in Eq.~(\ref{ops}), including constraints from recent LHC measurements of the Higgs decay rates, and from precision electroweak constraints, was studied in Ref.~\cite{Grojean:2013kd}. 

This paper constructs an exactly soluble model which generates the dimension six Higgs operators in Eq.~(\ref{ops}), with arbitrary coefficients consistent with the effective theory power counting. It also provides an explicit realization of the Lagrangian given in the appendix of Ref.~\cite{Jenkins:2013fya}.

The set of all dimension-six operators in the Standard Model was classified in Ref.~\cite{Grzadkowski:2010es}. There are 59 independent ones after redundant operators are eliminated by the equations of motion. The operators not involving fermions are the ones listed in Eq.~(\ref{ops}), their $CP$-odd partners $\widetilde O_{G},\widetilde O_{B},\widetilde O_{W},\widetilde O_{W\!B}$, four pure gauge operators of which two are $CP$ even and two are $CP$ odd,
\begin{align}
O_{G^3}&= f^{ABC} G^A_{\mu}{}^\nu G^B_{\nu}{}^\rho G^C_{\rho}{}^\mu,  
&\widetilde O_{G^3}&=  f^{ABC} \widetilde G^A_{\mu}{}^\nu G^B_{\nu}{}^\rho G^C_{\rho}{}^\mu, \nn  
O_{W^3}&= \epsilon^{abc}  W^a_{\mu}{}^\nu W^b_{\nu}{}^\rho W^c_{\rho}{}^\mu,   
&\widetilde O_{W^3}&=\epsilon^{abc} \widetilde W^a_{\mu}{}^\nu W^b_{\nu}{}^\rho W^c_{\rho}{}^\mu,
\end{align}
and three more operators involving the Higgs field,
\begin{align}
O_H &=\left(H^\dagger H\right)^3 ,\nn
O_{H \Box {}} &=\left(H^\dagger H\right) \partial^2 \left(H^\dagger H \right), \nn
O_{HD} & =\left(H^\dagger D_\mu H\right)^\dagger \left( H^\dagger D_\mu H\right)  \,.
\label{1.7}
\end{align}
The  exactly soluble model given here is a large-$N$ version of that  constructed in Ref.~\cite{Manohar:2006ga}. 
It produces the $CP$ conserving operators in Eq.~(\ref{ops}) with arbitrary order one coefficients, and the operator $O_{W^3}$, and does not generate any other dimension six operators at leading order in $1/N$.

Consider the Standard Model plus an additional scalar field $S^\alpha$ which is a weak $SU(2)$ doublet with hypercharge $Y_S$, and transforms as the $N$ dimensional representation of an internal $SU(N)$ global symmetry. $S_\alpha$ is a two component column vector, and $\alpha=1,\ldots, N$.
The theory is given by  
\begin{align}
\mathcal{L} &= \mathcal{L}_{\rm SM} + D_\mu S^{\dagger \alpha} D^\mu S_{\alpha} - V,
\end{align}
the usual Standard Model Lagrangian $L_{\rm SM}$, the $S_\alpha$ kinetic energy term, and the potential
\begin{align}
V &= m_S^2 S^{\dagger \alpha} S_{\alpha}  +
 \frac{\lambda_1}{N} H^{\dagger} H \, S^{\dagger \alpha} S_{\alpha} 
+ \frac{\lambda_2}{N} H^{\dagger}\tau^a H \, S^{\dagger \alpha} \tau^a S_{\alpha}\nn
& + \frac{\lambda_3}{N}  S^{\dagger \alpha} S_{ \alpha}\, S^{\dagger \beta} S_{\beta}   + \frac{\lambda_4}{N}  S^{\dagger \alpha}\tau^a S_{ \alpha}\, S^{\dagger \beta}\tau^a S_{\beta}
 \label{pot}
\end{align}
where the Standard Model Higgs potential $ \lambda\left(H^{\dagger} H- {v^2}/{2}\right)^2$ is part of $\mathcal{L}_{\rm SM}$. The Lagrangian is the most general renormalizable one consistent with the symmetries. Yukawa couplings of $S$ to the Standard Model fermions are forbidden by $SU(N)$ invariance. We assume that $m_S^2 >0$, so that $SU(N)$ is not spontaneously broken and the scalar mass $m_S$ is larger than the electroweak scale $v \sim 246$~GeV, so that the new interactions can be treated as higher dimension operators at the electroweak scale.

The large-$N$ limit of the theory is taken in the standard way~\cite{Coleman:1980nk}. One treats the theory in a perturbative expansion in the electroweak couplings $g \sim g_{1,2}$, i.e.\ one first expands in powers of $g$ and then takes the $N \to \infty$ limit. This is the usual method of computing weak decays in QCD using the $1/N$ expansion~\cite{Manohar:1998xv}.\footnote{The expansion has terms of order $(g^2 N)^r$, so we take the limit $g^2 \to 0$ first, followed by $N \to \infty$.  Equivalently, $N \gg 1$, and $g^2 N \ll 1$.} 

The method of Refs.~\cite{Coleman:1980nk,Coleman:1974jh,Abbott:1975bn} is used to solve the theory. Add to $V$ the dimension two auxiliary fields $\Phi$ and $\Psi^a$ which are real $SU(2)$ singlet and triplet fields with $Y=0$,
\begin{align}
V &\to V -\frac{\lambda_3 N}{4} \left( \frac{2}{N} S^{\dagger \alpha}S_{\alpha}+\frac{m_S^2}{\lambda_3} +\frac{\lambda_1}{\lambda_3N} H^\dagger H-\frac{\Phi}{\lambda_3} \right)^2\nn
& -\frac{\lambda_4 N}{4}  \left(  \frac{2}{N} S^{\dagger \alpha}\tau^a S_{\alpha}+\frac{\lambda_2}{\lambda_4N}  H^\dagger \tau^a H 
-\frac{\Psi^a}{\lambda_4} \right)^2
\label{npot}
\end{align}
The auxiliarly field equations of motion are
\begin{align}
\Phi &=m_S^2+\frac{\lambda_1}{ N} H^\dagger H ++ \frac{2 \lambda_3}{N}  S^{\dagger \alpha}S_{\alpha}\nn
\Psi^a &=   \frac{\lambda_2}{ N}  H^\dagger \tau^a H + \frac{2 \lambda_4 }{N} S^{\dagger \alpha}\tau^a S_{\alpha} 
\end{align}
which can be used to eliminate them and give back the original Lagrangian Eq.~(\ref{pot}).

In weak coupling, the scalar mass $m_S^2$ is  $\vev{\Phi}$. We will therefore use $\vev{\Phi}$ as the scale of new physics $\Lambda$ in Eq.~(\ref{h6}) and to normalize the operators in Eq.~(\ref{ops}).
Using $(H^\dagger \tau^a H)^2 = (H^\dagger H)^2$,
the new potential Eq.~(\ref{npot}) is
\begin{align}
V &=\frac{N}{2\lambda_3} m_S^2 \Phi-\frac{N}{4 \lambda_3} \Phi^2 - \frac{N}{4 \lambda_4} \Psi^a \Psi^a  + \frac{\lambda_1}{2\lambda_3} H^\dagger H \Phi \nn
& + \frac{\lambda_2}{2 \lambda_4} H^\dagger \tau^a H \Psi^a +\Phi S^{\dagger \alpha} S_\alpha +  \Psi^a S^{\dagger \alpha} \tau^a S_\alpha   -\frac{N m_S^4}{4\lambda_3} \nn
&-\frac{\lambda_1}{2\lambda_3}m_S^2 H^\dagger H  - \left( H^\dagger H\right)^2 \left(\frac{\lambda_1^2}{4 \lambda_3 N} +\frac{\lambda_2^2}{4 \lambda_4 N}\right)
\label{new}
\end{align}
The last term, which is subleading in $1/N$, as well as  the cosmological constant term, can be dropped.

The field $S_\alpha$ is now integrated out. This can be done exactly in the large-$N$ limit~\cite{Coleman:1980nk} to give an effective action which is an expansion in powers of $H$, $\Phi$ and $\Psi^a$. 
The Higgs field $H$ does not couple directly to $S$ in Eq.~(\ref{new}), so the $S$ functional integral generates terms which only depend on $\Phi$ and $\Psi^a$. The effective action has a derivative expansion in inverse powers of $m_S$, which will turn into a derivative expansion in inverse powers of $\vev{\Phi}$. The infrared divergences are controlled by $\vev{\Phi}$, since the theory is in the phase where the $SU(N)$ symmetry is unbroken and the $S$-sector is massive.

 At zero derivatives, one gets the Coleman-Weinberg effective potential~\cite{Coleman:1973jx} in the $\overline{\hbox{MS}}$ scheme
\begin{align}
V_{\rm CW} &= \frac{1}{64\pi^2}\text{Tr}\,  \left(M^2\right)^2 \left[-\frac32+\log \frac{M^2}{\mu^2}\right]
\end{align}
where
\begin{align}
\left[M^2\right]_{ab} &= \frac{\partial^2 V}{\partial \phi_a \partial \phi_b}
\end{align}
and $\phi_a=\text{Re}\,S_{\alpha,i}, \text{Im}\,S_{\alpha,i}$ are the scalar fields.
For the interaction in Eq.~(\ref{new}), $M^2$ has eigenvalues
$\Phi \pm \bm{\Psi}$, $\bm{\Psi}^2=\Psi^a \Psi^a$, each twice, 
so that
\begin{align}
V &= \frac{N}{2\lambda_3} \left(m_S^2+\frac{\lambda_1}{N}H^\dagger H\right) \Phi-\frac{N}{4 \lambda_3} \Phi^2 - \frac{N}{4 \lambda_4} \Psi^a \Psi^a   \nn
&+ \frac{\lambda_2}{2 \lambda_4} H^\dagger \tau^a H \Psi^a \nn
&+ \sum_{\pm} \frac{N}{32 \pi^2} \left(\Phi \pm \bm{\Psi}\right)^2\left[-\frac{3}{2}+\log \frac{ \Phi \pm \bm{\Psi} }{\mu^2}\right]
\label{new.d}
\end{align}
on integrating out the $S_\alpha$ field. This potential is exact in the large $N$ limit. $V$ is quadratic in $\bm{\Psi}$.  From the renormalization group (RG) equation for $V$, one finds that
\begin{align}
&\Phi, &&\Psi^a, &&\frac{1}{\lambda_3}\left(m_S^2+\frac{\lambda_1}{N}H^\dagger H\right), &&\frac{\lambda_2}{ \lambda_4} H^\dagger \tau^a H
\label{RGinv}
\end{align}
are $\mu$ independent\footnote{RG invariance refers to the dynamics of $S_\alpha$. The Standard Model fields are treated as external background fields.} and the exact  $\beta$-functions are
\begin{align}
\mu \frac{\rd \lambda_3}{\rd \mu} &= \frac{\lambda_3^2}{2\pi^2}, &\mu \frac{\rd \lambda_4}{\rd \mu} &= \frac{\lambda_4^2}{2\pi^2} \,.
\end{align}
Introduce the parameters $\Lambda_{3,4}$ in place of $\lambda_{3,4}(\mu)$,
\begin{align}
\frac{1}{\lambda_3(\mu)} &= \frac{1}{4\pi^2} \log \frac{\Lambda_3^2}{\mu^2}, 
&\frac{1}{\lambda_4(\mu)} &= \frac{1}{4\pi^2} \log \frac{\Lambda_4^2}{\mu^2},
\label{landau}
\end{align}
which are RG invariant. They are the scales at which $\lambda_{3,4}$ have a Landau pole, and at which the scalar theory breaks down. For consistency, we need $\Lambda_{3,4} > \vev{\Phi}$. New physics has to enter below $\Lambda_{3,4}$  for the theory to be valid to arbitarily high energies. For example $S_\alpha$ could be scalar fields generated by strong dynamics, or new interactions could enter which make the scalar couplings asymptotically free. We also define~\cite{Abbott:1975bn}
\begin{align}
\ks^2 &= \frac{m_S^2}{\lambda_3}
\label{kappa}
\end{align}
which is RG invariant, from Eq.~(\ref{RGinv}).

The effective action can be computed exactly in the large $N$ limit~\cite{Iliopoulos:1974ur,Chan:1985ny,Cheyette:1985ue,Zuk:1986xz,Chan:1986jq} in a derivative expansion. The terms which generate operators with dimension $\le 6$ in the Standard Model with coefficients that are non-vanishing in the $N \to \infty$ limit are
\begin{align}
\mathcal{L}_{S} &=   \frac{N}{96 \pi^2  \left(\Phi \right)}\left[  \partial_\mu \Phi \partial^\mu \Phi +  D_\mu \Psi^a D^\mu \Psi^a \right] \nn
& +\frac{N}{384\pi^2}  \left( \log \frac{\Phi}{\mu^2}\right)\biggl[  W_{\mu \nu}^a W^{a \mu \nu}
+4 Y_S^2 B_{\mu \nu}B^{\mu \nu}  \biggr] \nn
&+\frac{N}{192\pi^2 \left(\Phi\right)}  Y_S \Psi^a  W_{\mu \nu}^a B_{\mu \nu}
+ \mathcal{O}(g^3)
\label{l1}
\end{align}
where the gauge fields have been normalized so that the covariant derivative is $D_\mu = \partial_\mu + i W_\mu^a T_a + i B_\mu Y$. $\Phi$ is a gauge singlet, and has an ordinary derivative. $\Psi^a$ is in the $I=1$ representation of $SU(2)_W$ with $Y=0$, and has a covariant derivative
\begin{align}
D_\mu \Psi^a &= \partial_\mu \Psi^a + i  \left(T^c\right)_{ab} W^c_\mu \Psi^b,  &\left(T^c\right)_{ab} &= -i \epsilon_{cab}
\end{align}

\begin{figure*}
\begin{tabular}{ccccc}
\begin{minipage}{3.5cm} \includegraphics[width=3.5cm]{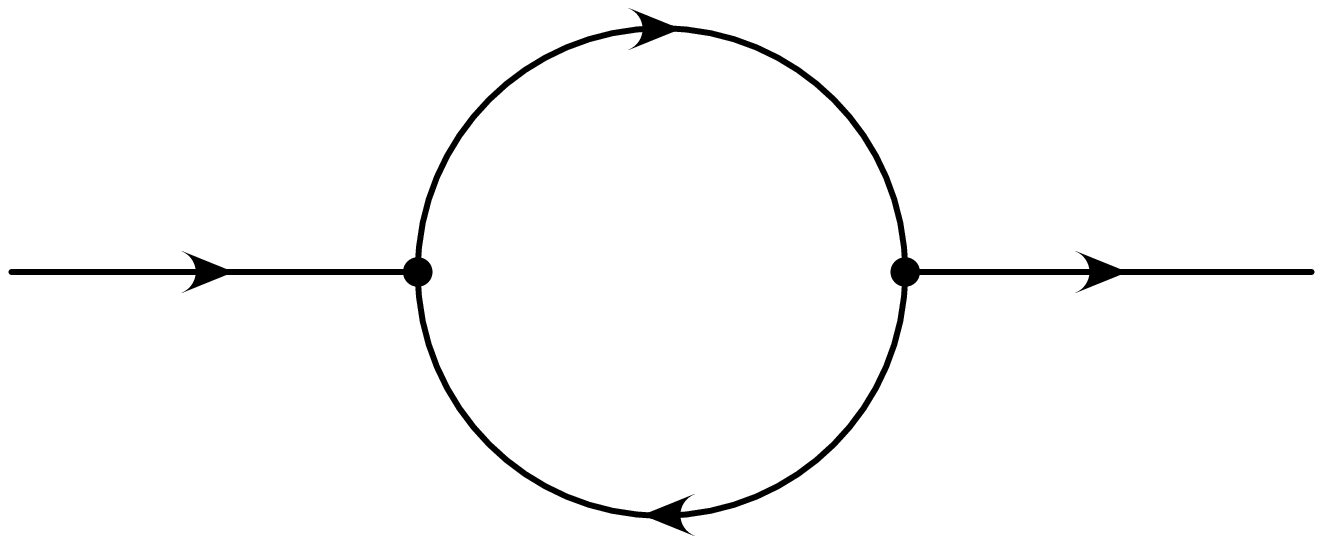}  \end{minipage} &
\begin{minipage}{3.5cm} \includegraphics[width=3.5cm]{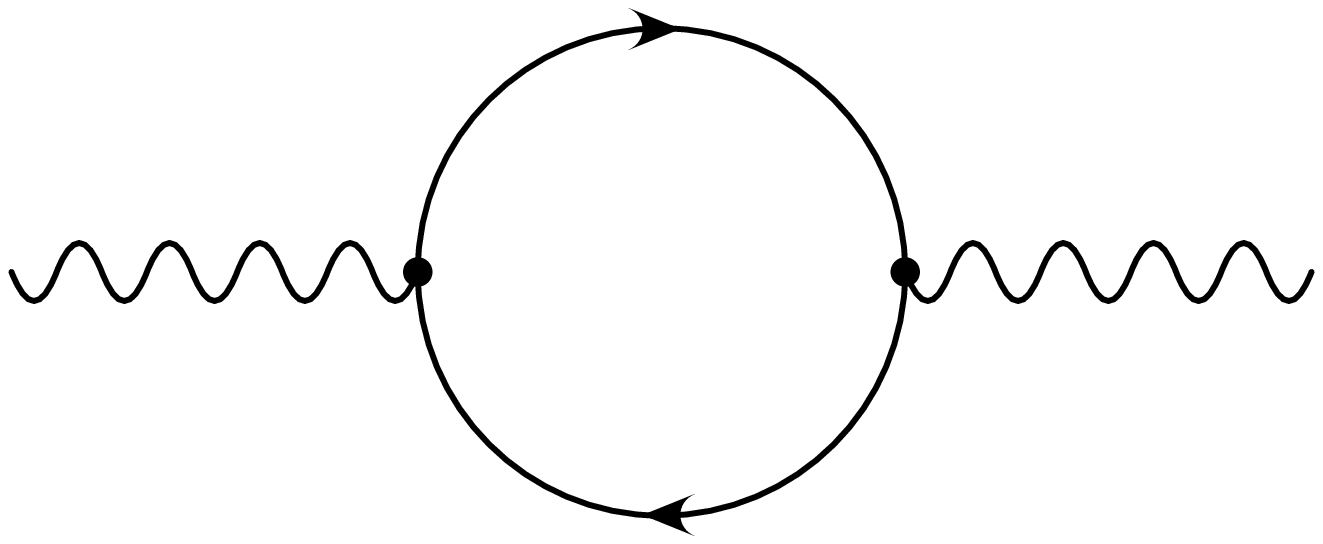}  \end{minipage} &
\begin{minipage}{3.5cm} \includegraphics[width=3.5cm]{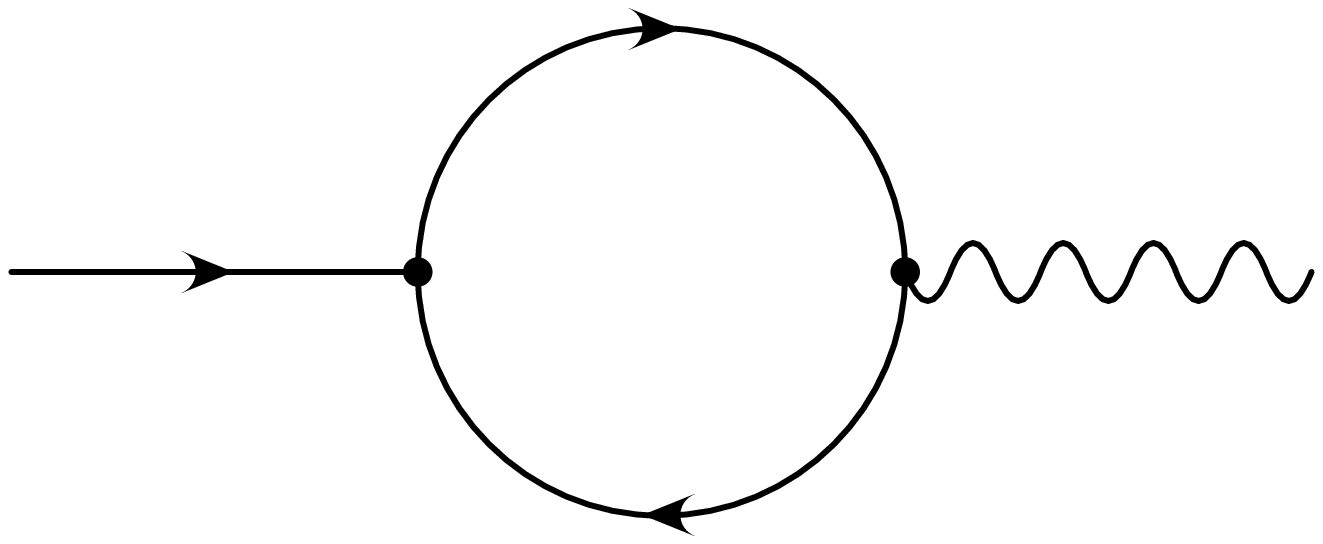}  \end{minipage} &
\begin{minipage}{3.5cm} \includegraphics[width=3.5cm]{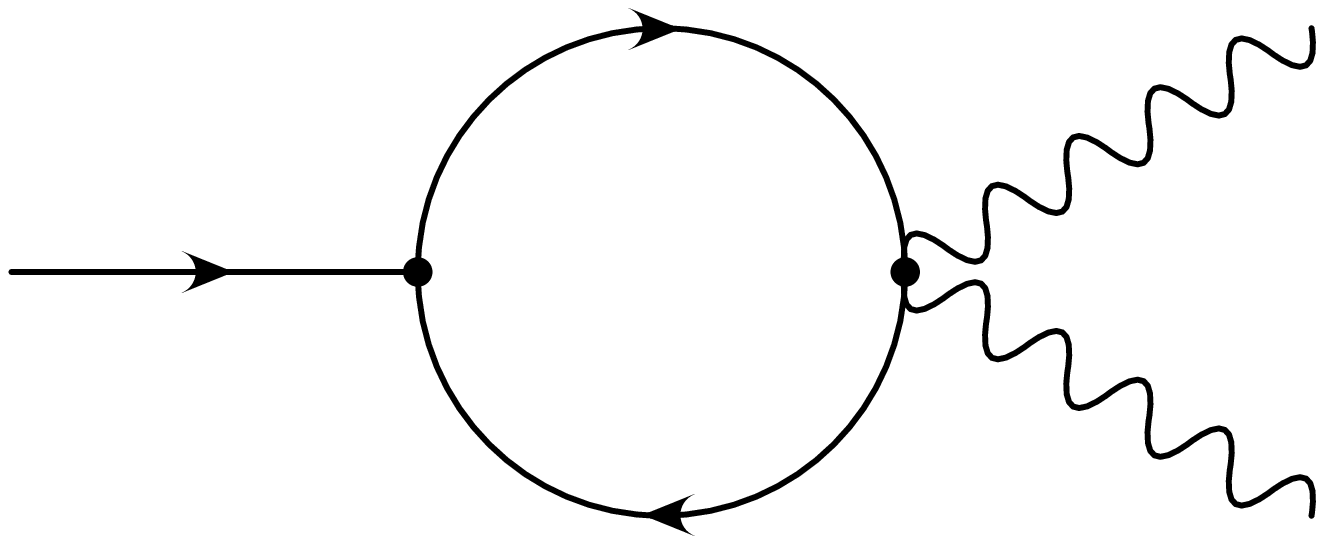}  \end{minipage} &
\end{tabular}

\begin{tabular}{cccccc}
\begin{minipage}{3.5cm} \includegraphics[width=3.5cm]{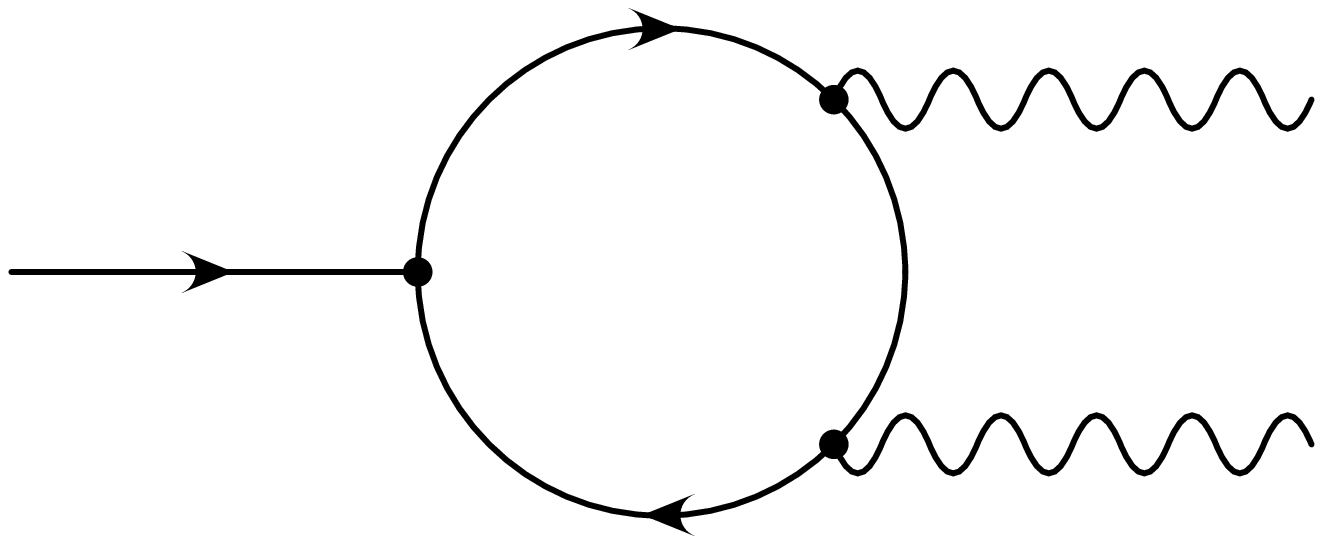}  \end{minipage} &
\begin{minipage}{3.6cm} \includegraphics[bb=115 194 497 599,width=3.5cm]{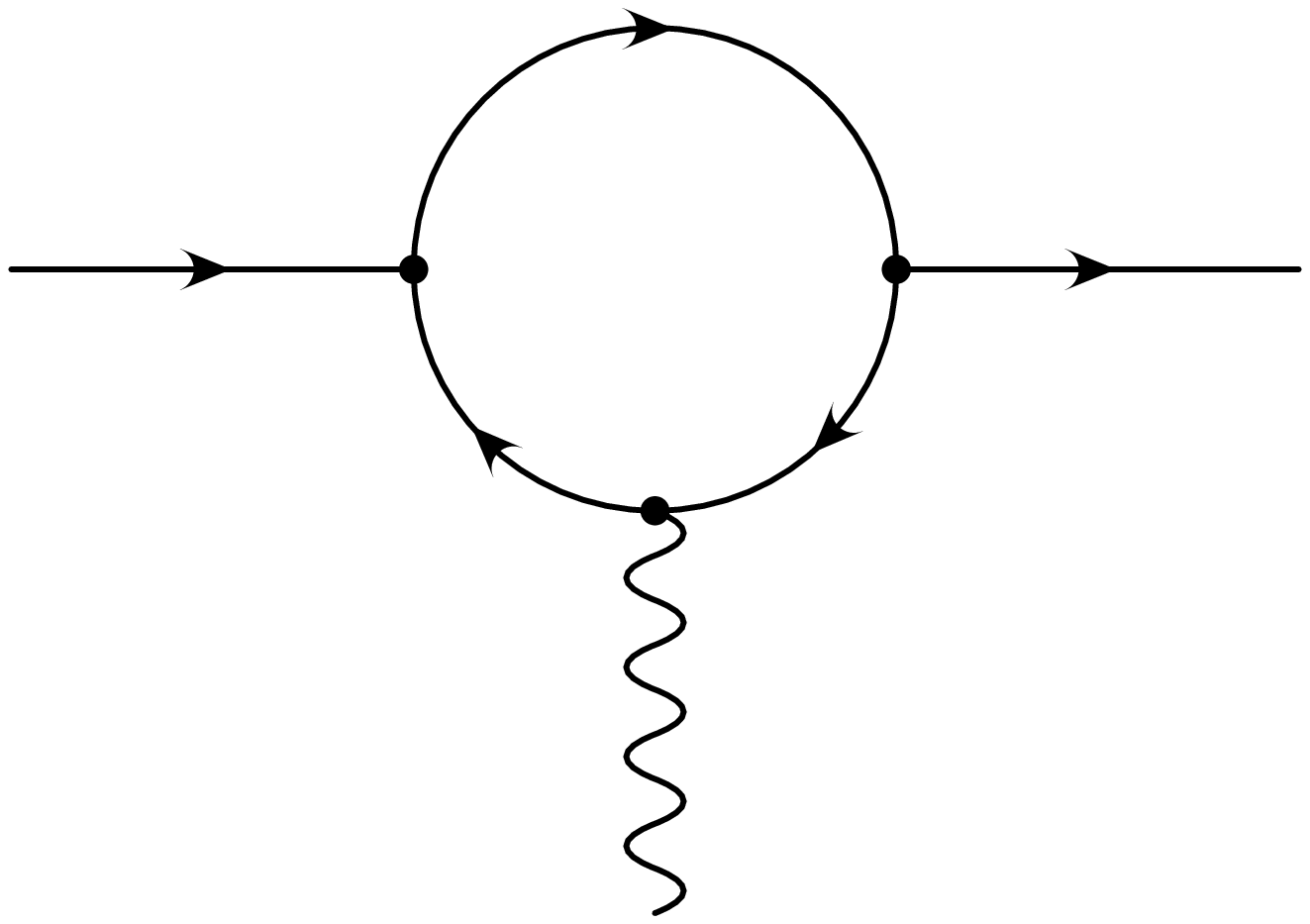}  \end{minipage} &
\begin{minipage}{3.6cm} \includegraphics[bb=115 237 497 559,width=3.5cm]{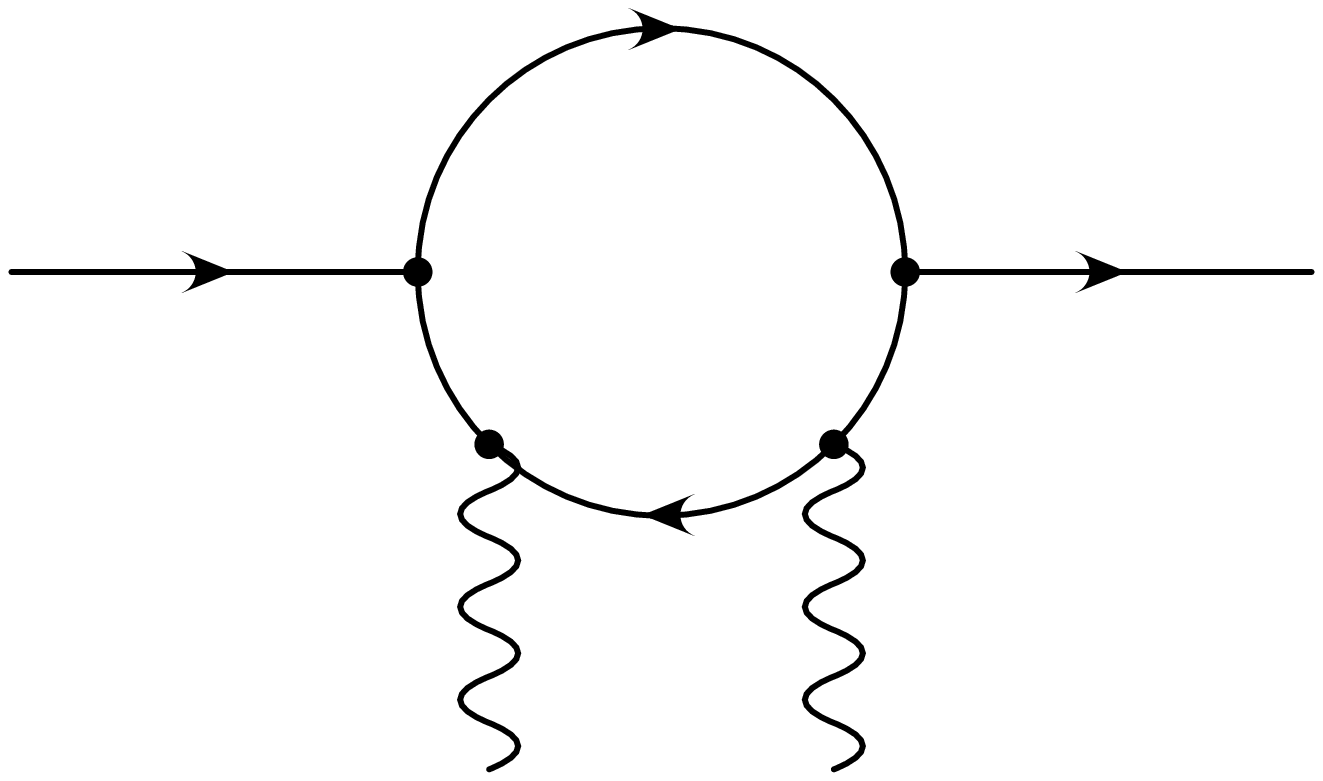}  \end{minipage} &
\begin{minipage}{3.6cm} \includegraphics[bb=115 194 497 599,width=3.5cm]{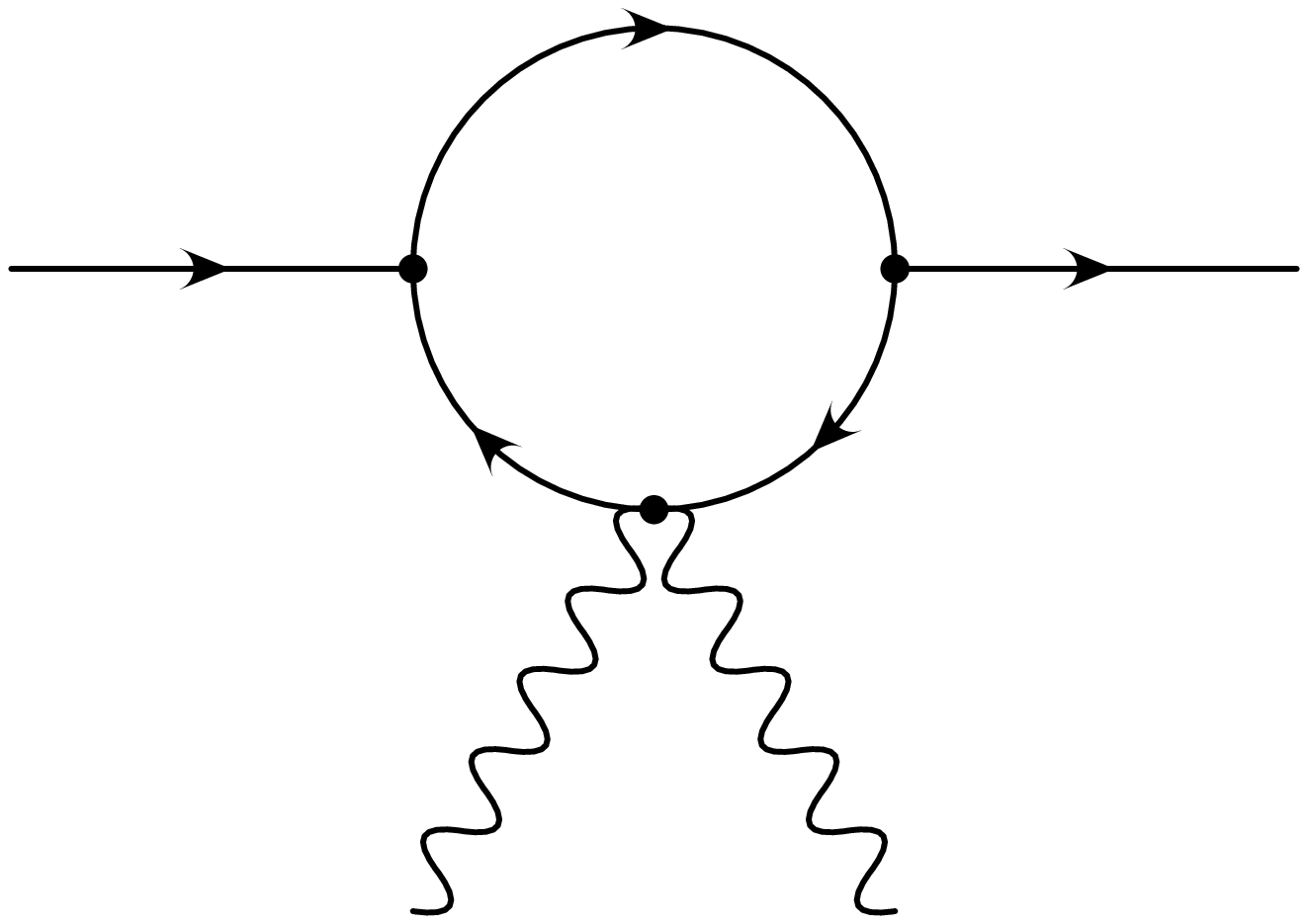}  \end{minipage} &
\end{tabular}

\begin{tabular}{cccccc}
\begin{minipage}{3.6cm} \includegraphics[bb=115 194 497 598,width=3.5cm]{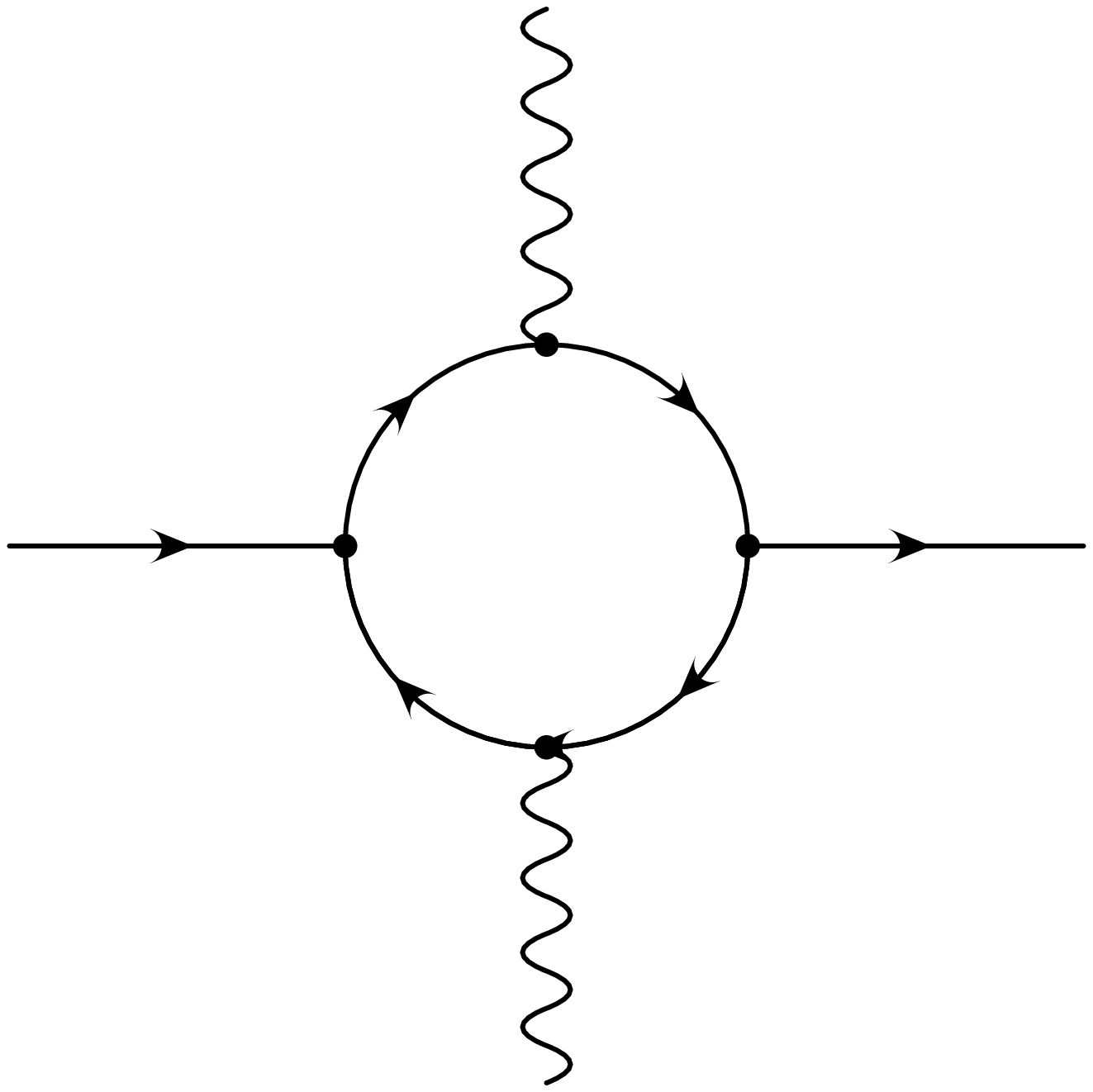}  \end{minipage} &
\begin{minipage}{3.6cm} \includegraphics[bb=115 194 497 598,width=3.5cm]{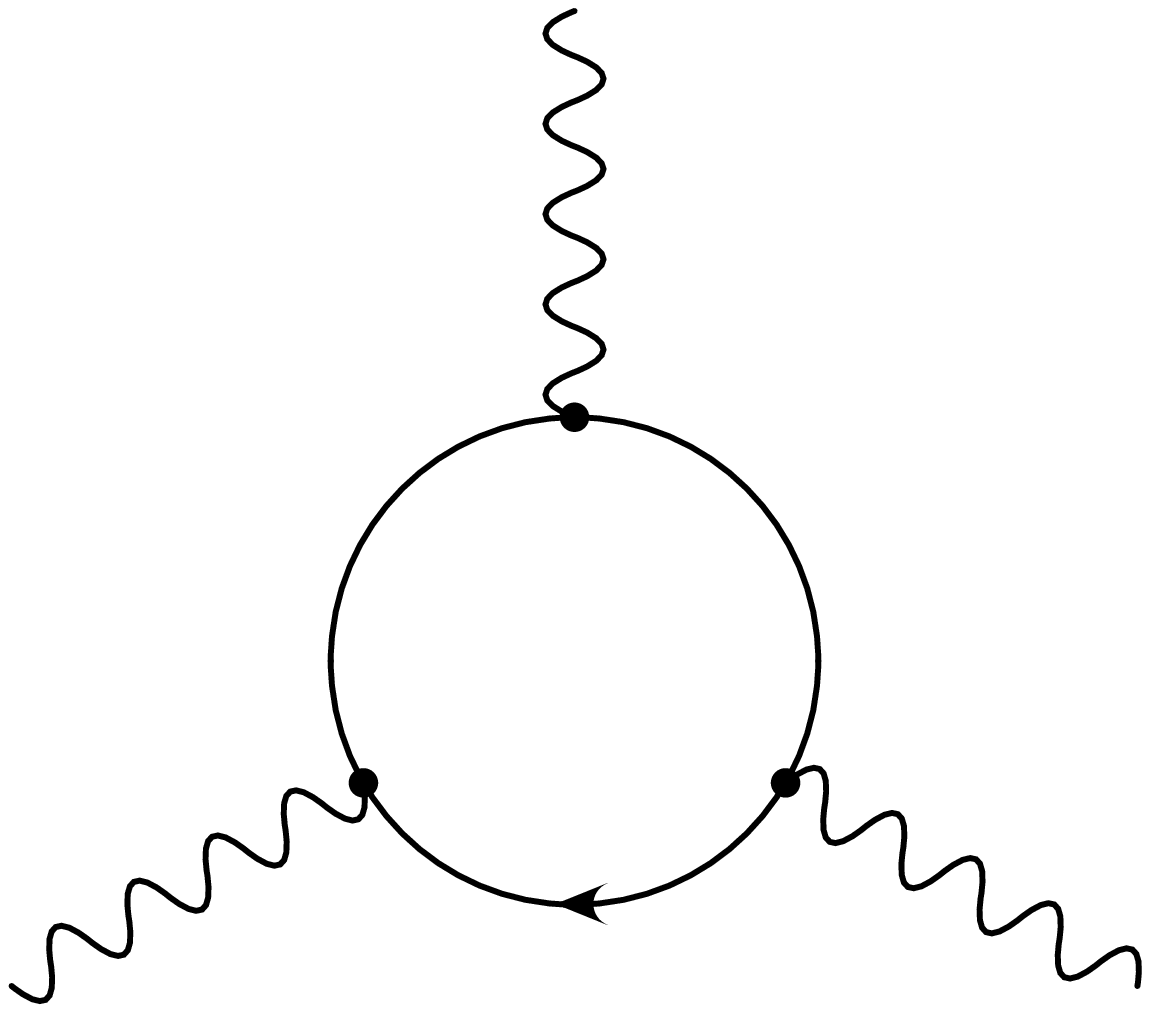}  \end{minipage}  &
\begin{minipage}{3.6cm} \includegraphics[bb=115 194 497 598,width=3.5cm]{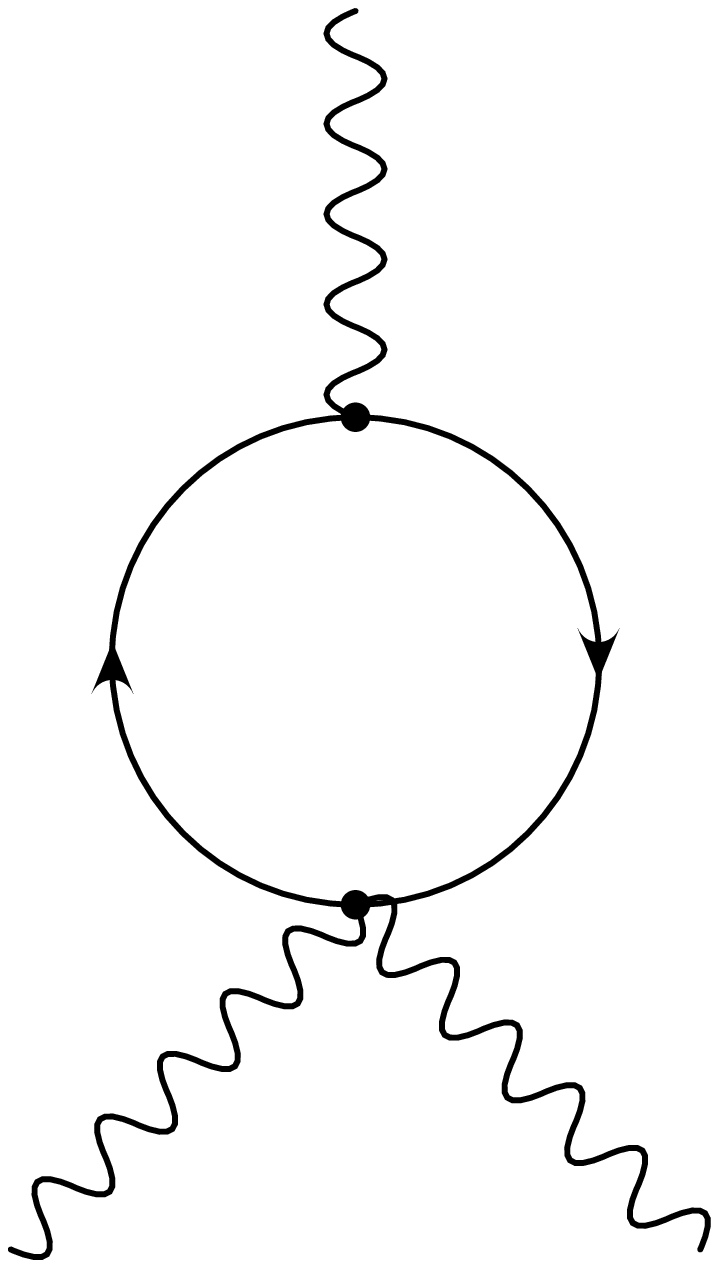}  \end{minipage} 
\end{tabular}
\caption{\label{fig}Graphs contributing to the dimension six effective action. The internal lines are $S_\alpha$ scalar fields. The external lines are $\Phi$, $\Psi^a$ and gauge fields.}
\end{figure*}

It is instructive to analyze Eq.~(\ref{l1}) at weak coupling. 
Let
\begin{align}
\Phi &= m_S^2 + \frac{4 \sqrt{3}\, \pi m_S}{ \sqrt{N}} \sigma, & \Psi^a &= \frac{4 \sqrt{3}\, \pi m_S}{ \sqrt{N}} \Sigma^a
\end{align}
The $\sigma$ and $\Sigma^a$ are dimension one fields with a canonically normalized kinetic energy term, and
\begin{align}
\mathcal{L}_{S} &= \frac12 \partial_\mu \sigma \partial^\mu \sigma + \frac12 D_\mu \Sigma^a D^\mu \Sigma^a 
\nn
& +\frac{N}{384\pi^2} \left( \log \frac{m_S^2}{\mu^2}\right) \biggl[ W_{\mu \nu}^a W^{a \mu \nu}
+4 Y_S^2 B_{\mu \nu}B^{\mu \nu}  \biggr] \nn
& +\frac{\sqrt{N}}{32 \sqrt{3}\,  \pi  m_S}\sigma \biggl[ W_{\mu \nu}^a W^{a \mu \nu}
+4 Y_S^2 B_{\mu \nu}B^{\mu \nu}  \biggr] \nn
&+\frac{\sqrt{N} Y_S }{16 \sqrt{3}\,  \pi  m_S}  \Sigma^a  W_{\mu \nu}^a B_{\mu \nu}
+ \mathcal{O}(g^3)\,.
\label{l2}
\end{align}
The effective action can be computed at weak coupling from the graphs in Fig.~\ref{fig}, which add up to give the gauge invariant structure in Eq.~(\ref{l2}).

The one-loop effective action generates kinetic energy terms for $\sigma,\Sigma^a$, given in the first line. The second line gives the threshold correction between the gauge couplings $g_h$ in the theory above $m_S$ and $g_l$ in the theory below $m_S$,
\begin{align}
 -\frac{1}{4 g_{l,2}^2(\mu)} &= -\frac{1}{4 g_{h,2}^2(\mu)}  + \frac{N}{384\pi^2} \log \frac{m_S^2}{\mu^2}, \nn
 -\frac{1}{4 g_{l,1}^2(\mu)}  &= -\frac{1}{4 g_{h,1}^2(\mu)}  + \frac{N}{384\pi^2} 4Y_S^2\log \frac{m_S^2}{\mu^2},
 \label{threshold}
\end{align}
for the $SU(2)$ and $U(1)$ coupling constants. The discontinuity in coupling matches the $S_\alpha$ contribution to the $\beta$-functions, which exists above $m_S$, but not below.  The remaining terms are $\sigma W_{\mu \nu}^a W^{a \mu \nu}$, 
$\sigma B_{\mu \nu}B^{\mu \nu}$ and $\Sigma^a W_{\mu \nu}^a B_{\mu \nu}$ interactions.

\begin{figure}
\includegraphics[width=4cm]{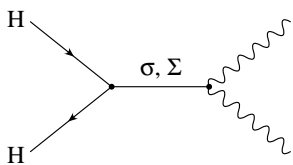}
\caption{\label{fig:ops} Graph generating the $h \to \gamma \gamma$ amplitude.}
\end{figure}

The scalar potential Eq.~(\ref{new.d}) becomes
\begin{align}
V &= -{12 \pi^2 m_S^2}\left( \frac{\sigma^2}{\lambda_3}+\frac{\Sigma^a \Sigma^a}{\lambda_4} \right) \nn
&+\frac{2 \sqrt {3}\, \pi}{\sqrt N} \left( \frac{\lambda_1}{\lambda_3} \sigma H^\dagger H + \frac{\lambda_2}{\lambda_4} \Sigma^a H^\dagger \tau^a H\right)
\end{align}
which has mass terms for $\sigma$ and $\Sigma^a$ (with the wrong sign, but it does not matter, they are auxiliary fields), and couplings of $\sigma$ and $\Sigma^a$ to the Higgs doublet. Integrating out the auxilary fields generates the operators Eq.~(\ref{ops})
 via the graph in Fig.~\ref{fig:ops}

We can now integrate out $\Phi,\Psi^a$ exactly, by doing the functional integral using the method of steepest descent~\cite{Iliopoulos:1974ur}.
The minimum $\vev{\Phi}$ is at
\begin{align}
\Phi \log \frac{\Phi}{\mu^2}-\Phi + \frac{4\pi^2}{\lambda_3} \left(m_S^2+\frac{\lambda_1}{N}H^\dagger H\right)
- \frac{4\pi^2}{\lambda_3} \Phi &=0.
\end{align}
Dropping the $1/N$ term, and using Eq.~(\ref{landau},\ref{kappa}), gives\footnote{$e=2.71828\ldots$.}
\begin{align}
\frac{\vev{\Phi}}{e \Lambda_3^2} \log \frac{\vev{\Phi}}{e \Lambda_3^2}+ \frac{4\pi^2\ks^2}{e \Lambda_3^2}&=0
\label{min}
\end{align}
in terms of RG invariant parameters. 
Instead of choosing $\ks$ as the Lagrangian parameter, and solving Eq.~(\ref{min}) for $\vev{\Phi}$, we can use $\vev{\Phi}$ as the free parameter and then determine $\ks$ from Eq.~(\ref{min}).  $\ks \to 0$ as  $\vev{\Phi} \to 0$. As $\vev{\Phi}$ increases, so does $\ks$, and $2 \pi \ks \to  \Lambda_3$ as $\vev{\Phi} \to \Lambda_3^2$. Eq.~(\ref{min}) implies that $\ks$ decreases again for $\vev{\Phi} > \Lambda_3$\cite{Coleman:1974jh,Abbott:1975bn}, but this is above the Landau pole, and the theory is not valid in this regime.

 Evaluating the functional integral around $\vev{\Phi}$ gives 
\begin{align}
\mathcal{L}_{S} &= \frac{N}{384\pi^2} \biggl[  \left( \log \frac{\vev{\Phi}}{\mu^2}\right) \left(W_{\mu \nu}^a W^{a \mu \nu}
+4 Y_S^2 B_{\mu \nu}B^{\mu \nu} \right) \biggr]\nn
& +\frac{\lambda_1}{2\lambda_3} \left(\lambda_3 \ks^2-\vev{\Phi}\right)H^\dagger H \nn
&+ \frac{\lambda_1}{96 \vev{\Phi} \lambda_3 \log \frac{\Lambda_3^2}{\vev{\Phi}}}  H^\dagger H \left(W_{\mu \nu}^a W^{a \mu \nu}
+4 Y_S^2 B_{\mu \nu}B^{\mu \nu} \right)  \nn
& +\frac{\lambda_2 Y_S}{48 \vev{\Phi} \lambda_4  \log \frac{\Lambda_4^2} {\vev{\Phi}}}  H^\dagger \tau^a H  W_{\mu \nu}^a B_{\mu \nu}\nn
&+\frac{N g_2^3}{2880 \pi^2 \vev{\Phi}}\epsilon^{abc}  W^a_{\mu}{}^\nu W^b_{\nu}{}^\rho W^c_{\rho}{}^\mu\,.
\label{ans1}
\end{align}
There are also terms at higher order in  the derivative expansion which have not been computed here. 
The first term in Eq.~(\ref{ans1}) gives the strong-coupling version  of the threshold correction Eq.~(\ref{threshold}). The second term is a shift in the Higgs mass proportional to the $S_\alpha$ mass, and can be absorbed into the $v^2$ term in the Higgs potential in $\mathcal{L}_{\rm SM}$. 

Using $\Lambda=\vev{\Phi} \sim m_S^2 > v$ as the scale in Eq.~(\ref{ops}), we see that we have generated the Standard Model Lagrangian plus the three $CP$-even dimension six operators in Eq.~(\ref{h6}) with coefficients
\begin{align}
c_W &= \frac{(\lambda_1/\lambda_3)}{48 \log \frac{\Lambda_3^2}{\vev{\Phi}}}\,, \nn
c_B &= \frac{(\lambda_1/\lambda_3) Y_S^2 }{12   \log \frac{\Lambda_3^2}{\vev{\Phi}}}\,,\nn
c_{W\!B} &= \frac{(\lambda_2/\lambda_4)Y_S}{24   \log \frac{\Lambda_4^2}{\vev{\Phi}}}\,,
\label{result}
\end{align}
and the $O_{W^3}$ operator with coefficient
\begin{align}
c_{W^3} &=\frac{N g_2^3}{2880 \pi^2 \vev{\Phi}}\,.
\end{align}
All other dimension six operators are subleading in $1/N$. 
The ratios $(\lambda_1/\lambda_3)$ and $(\lambda_2/\lambda_4)$ are RG invariant under $S_\alpha$ dynamics with the Standard Model fields treated as background fields, from Eq.~(\ref{RGinv}).

The linear combinations of coefficients relevant for $h \to \gamma \gamma$ and $h \to \gamma Z$ decays are
\begin{align}
c_{\gamma \gamma} &=c_{W}+c_{B} -c_{W\!B}\,, \nn
c_{\gamma Z} &= c_{W} \cot \theta_W -c_{B} \tan \theta_W -c_{W\!B} \cot 2 \theta_W\,.
\end{align}
The operator $c_{W\!B}$ is constrained by the $S$-parameter~\cite{Grinstein:1991cd, Hagiwara:1993ck, Alam:1997nk, Han:2004az},
\begin{align}
S &= - {8 \pi^2 v^2} \frac{c_{W\!B}}{\Lambda^2}\,.
\end{align}
From Eq.~(\ref{result}), we see that we can get order unity values of $c_W$, $c_B$ and $c_{W\!B}$. The phenomenology of the Higgs-gauge operators was discussed in detail in Refs.~\cite{Manohar:2006gz,Manohar:2006ga,Grojean:2013kd}.

There is one relation that follows from Eq.~(\ref{result}),
\begin{align}
c_B &= 4 Y_S^2 c_W\,,
\end{align}
if we restrict to the model considered here with a single scalar multiplet with hypercharge $Y_S$.
One can construct trivial generalizations of the large $N$ model with multiple heavy scalar fields, which can have different hypercharges, and can also be colored. In this case, one can also generate the gluon term $c_G$, as in the octet scalar model of Ref.~\cite{Manohar:2006ga}, and the $c_B-c_W$ relation no longer holds. 

The large $N$ calculation drops terms of order $1/N$, as well as higher order radiative corrections of order $g_2^2 N/(16\pi^2)$.
For finite $N$, the neglected terms are small if $1 \ll N \ll 400$. It would be interesting to explore the full parameter space of scalar couplings and masses where the potential is stable and $m_S$ is below the Landau pole, to determine the allowed region for $c_W$, $c_B$ and $c_{W\!B}$. 

The $S_\alpha$ interactions break custodial $SU(2)$ symmetry, since $S_\alpha$ is in a complex  representation of $SU(N)$, and the real and imaginary parts of $S_\alpha$ cannot be combined to form an $O(4)$ vector, as is possible for the Higgs field. 
This does not affect the standard relations such as $M_W=M_Z \cos \theta_W$ that follow from custodial $SU(2)$ symmetry in the Higgs sector. Custodial $SU(2)$ symmetry violation due to $S_\alpha$ interactions only arise from higher dimension operators.
One can also study variants of the theory with $SO(N)$ symmetry, or double the $S_\alpha$ fields to have a $O(4) \times SU(N)$ symmetry. In these variants, custodial $SU(2)$ can be incorporated in the $S$ potential.

The model has threshold corrections to the gauge couplings, Eq.~(\ref{threshold}), which affects gauge unification. This was studied in Ref.~\cite{Manohar:2006gz}. In general, all theories that introduce new dynamics will modify the standard unification scenario, which has perturbative unification with a desert up to the GUT scale.

Finally, the model needs a fine tuning of order $1\%$ to keep $m_H$ small compared to the scale $\Lambda \sim m_S$, since there is contribution to $m_H^2 \propto m_S^2$ in Eq.~(\ref{ans1}). While not desirable, this is not worse than  fine-tunings required in many models proposed to solve the hierarchy problem. The $H$ mass term and dimension six operators have different dependence on the RG invariant parameters, so theories with additional $S$ multiplets can cancel the $m_H$ contribution without cancelling the Higgs-gauge operators, if the parameters satisfy
\begin{align}
\sum_i \lambda_{1,i} {\ks}_{,i}^2-\frac{\lambda_{1,i}}{\lambda_{3,i}} \vev{\Phi_i} &=0\,.
\label{31}
\end{align}
This cancellation condition is not adjusted order-by-order in perturbation theory, since Eqs.~(\ref{ans1},\ref{31}) are exact at leading order in $1/N$.  The Higgs mass is then light because it is $1/N$ suppressed.

I would like to thank E.~Jenkins and M.~Trott for helpful discussions.

\newpage

\bibliography{paperrefs}

\end{document}